\documentclass[prc,letterpaper,twocolumn,showpacs,showkeys,lengthcheck,tightenlines,
               nofootinbib,preprintnumbers,superscriptaddress]{revtex4-1}

\usepackage[utf8]{inputenc}

\usepackage{titlesec}
\usepackage{dcolumn}
\usepackage{bm}
\usepackage{amsmath,amssymb,amsfonts}

\usepackage{graphicx}
\usepackage{subfigure}
\usepackage{color}
\usepackage{bbold}

\usepackage{slashed}
\usepackage[svgnames]{xcolor}
\usepackage[english]{babel}
\usepackage{blindtext}
\usepackage{microtype}
\usepackage{tikz}

\usepackage[colorlinks=true,urlcolor=blue,citecolor=blue]{hyperref}
\usepackage{ifpdf}
\usepackage{float}

\graphicspath{{.}{figures/}}

\titlespacing{\section}{5pt}{12pt plus 4pt minus 2pt}{8pt plus 2pt minus 2pt}
\titlespacing{\subsection}{0pt}{12pt plus 4pt minus 2pt}{8pt plus 2pt minus 2pt}

\begin{document}

\title{Pion-photon and kaon-photon transition distribution amplitudes in the Nambu--Jona-Lasinio model}

\author{Jin-Li Zhang}
\email[]{jlzhang@njit.edu.cn}
\affiliation{Department of Mathematics and Physics, Nanjing Institute of Technology, Nanjing 211167, China }

\author{Jun Wu}
\email[]{wujun@njit.edu.cn}
\affiliation{Department of Mathematics and Physics, Nanjing Institute of Technology, Nanjing 211167, China }
\affiliation{National Laboratory of Solid State Microstructures, Department of Physics, Nanjing University, Nanjing 210093, China }





\begin{abstract}
The Nambu--Jona-Lasinio model is utilized to investigate the pion and kaon photon leading-twist transition distribution amplitudes using proper time regularization.
Separately, the properties of the vector and axial vector pion photon transition distribution amplitudes are examined, and the results meet the desired properties. Our study involves sum rule and polynomiality condition. The vector and axial vector pion photon transition form factors that are present in the $\pi^+\rightarrow \gamma e^+ \nu$ process are the first Mellin moments of the pion photon transition distribution amplitudes. The vector transition form factor originates from the internal structure of hadrons, the axial current can be coupled to a pion, this pion is virtual, and its contribution will be present independently of the external hadrons. The kaon transition form factors are similar. The vector form factor's value at zero momentum transfer is determined by the axial anomaly, while this is not the case for the axial one. The vector and axial form factors, as well as the neutral pion vector form factor $F_{\pi \gamma \gamma}(t)$, are depicted. According to our findings, the pion axial transition form factor is harder than the vector transition form factor and harder than the electromagnetic form factor. We also discuss the link between $\pi - \gamma $ and $\gamma - \pi$ transitions distribution amplitudes.
\end{abstract}


\maketitle
\section{Introduction}
Hard exclusive processes are an important tool for exploring the internal quark-gluon dynamics of hadrons. Off-diagonal exclusive processes in perturbative Quantum Chromodynamics (QCD) are the subject of numerous experimental and theoretical works. In high-energy physics, factorization theorems are a tool that can be used to separate perturbatively computable parts from non-perturbative matrix elements in a systematic way. By using factorization theorems, one can depict the amplitudes of these reactions as the combination of a hard scattering kernel and a soft matrix element of quark and/or gluon fields. The soft matrix element relates to the generalised parton distributions (GPDs)~\cite{Mueller:1998fv,Ji:1996nm,Radyushkin:1997ki,Ji:1998pc,Belitsky:2001ns,Goeke:2001tz,Wang:2001ifa,Diehl:2002he,Burkardt:2002hr,Diehl:2003ny,Belitsky:2005qn,Boer:2011fh,Zhang:2020ecj,Zhang:2021mtn,Zhang:2021shm,Zhang:2021tnr,Zhang:2021uak,Zhang:2022zim,Chavez:2021llq,Raya:2021zrz,Adhikari:2021jrh}. Deeply virtual Compton scattering (DVCS) is a typical example of processes related to GPDs. A generalization of GPDs to non-diagonal transitions was firstly introduced through hadron-photon transitions in Refs.~\cite{Pire:2004ie,Pire:2005ax,Bianconi:2004wu}, which are the transition distribution amplitudes (TDAs)~\cite{Tiburzi:2005nj,Dorokhov:2006qm,Courtoy:2008ij,Kotko:2009ij,Bakulev:2011rp}. TDAs relate to the situation where the initial and final states correspond to distinct particles. The description of exclusive high-energy processes~\cite{lepage1980exclusive,brodsky1981helicity,brodsky1999semi,Melic:2001wb,braun2008exclusive,radyushkin2009shape,polyakov2009pion,Dumm:2013zoa,radyushkin2017pion} requires the presence of the fundamental theoretical ingredient called pion distribution amplitude ($\pi$DA). TDAs appear as the non-perturbative quantities in the analysis of the virtual Compton scattering and other exclusive reactions, such as the hadron-anti-hadron annihilation process, $H\bar{H}\rightarrow \gamma\gamma^*$. Vector and axial pion-to-photon TDAs are required as a non-perturbative input to investigate the cross-section of pion annihilation into the virtual and real photon. In some sense, TDAs are a mixture of the ordinary PDAs and GPDs~\cite{Kotko:2009ij,Belitsky:2005qn}. Pion photon TDA reduce to the ordinary PDA in the forward limit $t=0$ and $\xi=0$. Unlike the GPDs, TDAs are defined as hadron-photon matrix elements of non-local operators. This means that TDAs concern not only the momenta, but also the physical states, so TDAs are not time reversal invariant. TDAs should satisfy the symmetries required by the QCD, such as the sum rule. This means the Mellin moments of TDAs correspond to the transition form factors~\cite{CLEO:1997fho,Feldmann:1997vc,Kurimoto:2001zj,BaBar:2009rrj,Balakireva:2012rq,Raya:2015gva,Raya:2016yuj,Redmer:2019iye,Stefanis:2019cfn,Stefanis:2020rnd}. Mellin moments are related to the vector and axial vector transition form factors. A comparison is made between the two form factors and electromagnetic form factors (FFs)~\cite{Xu:2019ilh,Cui:2020rmu}. Moreover, they should satisfy the polynomiality condition, so that the coefficients of the higher Mellin moments of TDAs can be studied.

In the experimental side, there is little known about GPDs, especially TDAs (see a review~\cite{Pire:2011jh}). The situation is somewhat better on the theory side, especially for the processes of simple hadronic states. It is easiest to consider pion-photon TDAs~\cite{Tiburzi:2005nj,Courtoy:2007vy,Broniowski:2007fs,Courtoy:2008af,Courtoy:2008ij,Kotko:2008gy,Kotko:2009ij,Kotko:2013ima,Kotko:2012cna}, the nonperturbative feature of the distribution functions lead to the use of effective models or phenomenological parametrization. This paper is not only interested in the pion-photon TDAs in the Nambu--Jona-Lasinio (NJL)~\cite{RevModPhys.64.649,Buballa:2003qv} model, but also kaon photon TDAs. The TDAs of pion photon are important as a nonperturbative input in the cross-section of pion annihilation into real and virtual photons. The study of pion photon TDAs is conducted in non-local chiral quark models~\cite{Kotko:2009ij,Kotko:2012cna,Kotko:2013ima}, including the NJL model, which employs the Pauli-Villars regularization scheme~\cite{Courtoy:2007vy,Courtoy:2008pb,Courtoy:2008af}, as well as in simple analytical models~\cite{Tiburzi:2005nj}, including the spectral quark model~\cite{Broniowski:2007fs}. The NJL model is widely used in many fields because it has an effective Lagrangian of relativistic fermions that interact through local fermion-fermion couplings. It is noteworthy that the NJL model preserves the fundamental symmetries of QCD. Among these symmetries, the chiral symmetry holds significant importance. It is important to know that pions are Gold-stone bosons with broken SU(2) chiral symmetry, and their properties are heavily depend on the symmetry breaking. It is possible to gain insight into the pion and kaon photon TDAs using the NJL model. NJL model is non-renormalizable theory and a cutoff procedure is needed to completely define the model, the proper time regularization scheme is employed in this paper.

This paper is organized as follows: In Sec.~\ref{nice}, A brief introduction to NJL model is given. We give the definition and calculation of pion photon and kaon photon TDAs, in addition, basic properties of TDAs will be checked in this section.  A brief summary and discussions are given in Sec.~\ref{excellent}.


\section{TDAs}\label{nice}

\subsection{NJL model}\label{good}
The SU(2) flavor NJL Lagrangian is defined as
\begin{align}\label{1}
\mathcal{L}=\bar{\psi }\left(i\gamma ^{\mu }\partial _{\mu }-\hat{m}\right)\psi+G_{\pi }\left[\left(\bar{\psi }\lambda_a\psi\right)^2-\left( \bar{\psi }\gamma _5 \lambda_a \psi \right)^2\right],
\end{align}
where $\vec{\tau}$ are the Pauli matrices representing isospin and $\hat{m}=\text{diag}\left(m_u,m_d,m_s\right)$ is the current quark mass matrix. In the limit of exact isospin symmetry, $m_u = m_d =m$, where the quark field has the flavor components $\psi^T = (u, d, s)$, $\lambda_a$, $a=0,\ldots,8$ are the eight Gell-Mann matrices in flavor space where $\lambda_0=\sqrt{2/3}$ $\mathbb{1}$. $G_{\pi}$ is an effective coupling strength of the scalar ($\bar{q}q$) and pseudoscalar ($\bar{q}\gamma_5 q$) interaction channels.

The NJL dressed quark propagator is obtained by solving the gap equation, the solution is
\begin{align}\label{2}
S_q(k)=\frac{1}{{\not\!k}-M_q+i \varepsilon}.
\end{align}
The interaction kernel of the gap equation is local, so we obtain a constant dressed quark mass $M$ which satisfies
\begin{align}\label{3}
M_q=m_q+12 i G_{\pi}\int \frac{\mathrm{d}^4l}{(2 \pi )^4}\text{Tr}_D[S_q(l)],
\end{align}
where the trace is over Dirac indices. Only for a strong coupling $G_{\pi} > G_{critical}$ can dynamical chiral symmetry breaking (DCSB) happen, which gives a nontrivial solution $M_q > 0$.

The NJL model is a non-renormalizable quantum field theory. A regularization method must be used to fully define the model. We will use the proper time regularization scheme~\cite{Ebert:1996vx,Hellstern:1997nv,Bentz:2001vc}.
\begin{align}\label{4}
\frac{1}{X^n}&=\frac{1}{(n-1)!}\int_0^{\infty}\mathrm{d}\tau \tau^{n-1}e^{-\tau X}\nonumber\\
& \rightarrow \frac{1}{(n-1)!} \int_{1/\Lambda_{\text{UV}}^2}^{1/\Lambda_{\text{IR}}^2}\mathrm{d}\tau \tau^{n-1}e^{-\tau X}
\end{align}
where $X$ represents a product of propagators that have been combined using Feynman parametrization.

In the NJL model, the description of a meson as a $\bar{q}q$ bound state is obtained through the Bethe-Salpeter equation (BSE). The BSE in each meson channel is given by a two-body $t$-matrix that depends on the type of interaction channel. The reduced $t$-matrices for the pion and kaon meson reads,
\begin{align}\label{3}
\tau_{\pi,K}(q)=\frac{-2i\,G_{\pi}}{1+2G_{\pi}\Pi_{PP}^{\pi,K}(q^2)},
\end{align}
where the bubble diagram $\Pi_{PP}(q^2)$ are defined as
\begin{align}\label{ppbb}
\Pi_{PP}^{\pi,K}(q^2)\delta_{ij}=3i\int \frac{d^4k}{(2\pi)^4}\text{Tr}[\gamma^5\tau_iS_u(k)\gamma^5\tau_jS_{d,s}(k+q)],
\end{align}
where the traces are over Dirac and isospin indices.
%
%

The mass of pion and kaon are given by the poles in the reduced $t$-matrix,
\begin{align}\label{5}
1+2\,G_{\pi}\Pi_{PP}^{\pi,K}(q^2=m_{\pi}^2,m_K^2)=0.
\end{align}
Expanding the full $t$-matrix about the pole gives the homogeneous Bethe-Salpeter vertex for the pion and kaon,
\begin{align}\label{6C}
\Gamma_{\pi,K}^{i}=\sqrt{Z_{\pi,K}}\gamma_5\lambda_i,
\end{align}
the normalization factor is given by
\begin{align}\label{qmcc}
Z_{\pi,K}^{-1}&=-\frac{\partial}{\partial q^2}\Pi_{PP}^{\pi,K}(q^2)|_{q^2=m_{\pi}^2,m_K^2}.
\end{align}
These residues can be explained as the square of the effective meson-quark-quark coupling constant. Homogeneous Bethe-Salpeter vertex functions are an essential ingredient, for example, in triangle diagrams that determine the meson form factors.

In addition to the ultraviolet cutoff, $\Lambda_{\text{UV}}$, we also include the infrared cutoff $\Lambda_{\text{IR}}$. The NJL model doesn't contain confinement, the infrared cutoff is used to mimic confinement. The parameters used in this work are given by Table \ref{tb1}, and $f_{\pi}=0.092$ GeV, $f_{K}=0.092$ GeV.

\begin{center}
\begin{table}
\caption{Parameter set used in our work. The dressed quark mass and regularization parameters are in units of GeV, while coupling constant are in units of GeV$^{-2}$.}\label{tb1}
\begin{tabular}{p{0.8cm} p{0.8cm} p{0.8cm} p{0.8cm}p{0.8cm}p{0.8cm}p{0.8cm}p{0.9cm}p{0.9cm}}
\hline\hline
$\Lambda_{\text{IR}}$&$\Lambda _{\text{UV}}$&$M_u$&$M_s$&$G_{\pi}$&$m_{\pi}$&$m_K$&$Z_{\pi}$&$Z_K$\\
\hline
0.240&0.645&0.4&0.59&19.0&0.14&0.47&17.85&20.47\\
\hline\hline
\end{tabular}
\end{table}
\end{center}

\subsection{The definition and calculation of pion photon TDAs}\label{qq}
The pion photon TDAs in the NJL model, $p$ is the incoming pion momentum, $p'$ is the outgoing photon momentum, and $\varepsilon$ is the polarization. The symmetry notation as GPDs will be used in this paper, the alignment of notation used in the manuscript with that employed in Ref.~\cite{Courtoy:2007vy}, the kinematics of this process and the relevant quantities will be defined as
\begin{align}\label{4}
p^2=m_{\pi}^2,\quad p^{'2}=0, \quad q^2=\Delta^2=(p'-p)^2=t,
\end{align}
\begin{align}\label{5}
\xi=\frac{p^+-p'^+}{p^++p'^+}, \quad n^2=0,
\end{align}
$\xi$ is the skewness parameter, in the light-cone coordinate
\begin{align}\label{4A}
v^{\pm}=(v^0\pm v^3), \quad  \mathbf{v}=(v^1,v^2)
\end{align}
for any four-vector, $n$ is the light-cone four-vector defined as $n=(1,0,0,-1)$, then the $v^+$ in the light-cone coordinate can be represented as
\begin{align}\label{4B}
v^+=v\cdot n.
\end{align}
\begin{figure}
\centering
\includegraphics[width=0.2\textwidth]{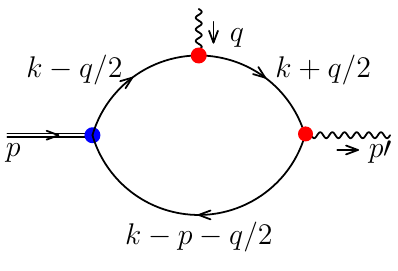}
\qquad
\includegraphics[width=0.2\textwidth]{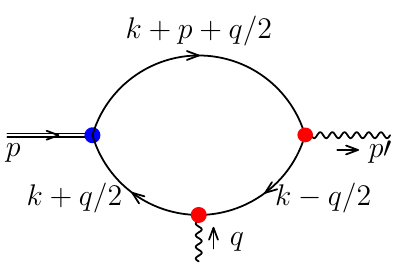}
\caption{The $\pi^+$ photon TDAs. The double lines represent $\pi^+$, the single line corresponds to the quark propagator, the wavy line corresponds to the photon, the blue dot represents the $\bar{q}q$ interaction kernel, the red dots represent the quark photon vertex. The left panel represents a quark $u$ changed into a quark $d$ by the bilocal current, the right panel represents a quark $d$ changed into a quark $u$. }\label{piontda}
\end{figure}

\begin{figure}
\centering
\includegraphics[width=0.30\textwidth]{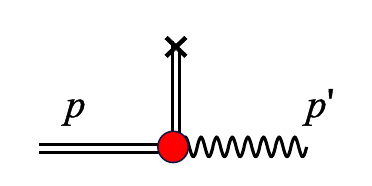}
\caption{Pion pole contribution between the axial current (the cross) and the incoming pion-photon vertex, the double line represent the pion, the wavy line corresponds to the photon, the red dot represents the pion-photon vertex. }\label{pionpole}
\end{figure}

TDAs are defined as the Fourier transform of the matrix element of bilocal currents at a light-like distance. The definition of the two leading-twist TDAs are,
\begin{widetext}
\begin{align}\label{vdtda}
\int \frac{\mathrm{d}z^-}{2\pi}e^{i xP^+z^-} \langle \gamma(p')|\bar{q}(-\frac{z}{2})\gamma^+ q(\frac{z}{2})|\pi^+(p)\rangle \mid_{z^+=0,\bm{z}=\bm{0}}=i e \varepsilon^{\nu} \epsilon^{+\nu\rho\sigma} P^{\rho}\Delta^{\sigma} \frac{V^{\pi^+}(x,\xi,t)}{f_{\pi}},
\end{align}
\begin{align}\label{atda}
\int \frac{\mathrm{d}z^-}{2\pi}e^{ixP^+z^-} \langle \gamma(p')|\bar{q}(-\frac{z}{2})\gamma^+\gamma^5q(\frac{z}{2})|\pi^+(p)\rangle \mid_{z^+=0,\bm{z}=\bm{0}}&=e (\vec{\varepsilon}^{\perp}\cdot \vec{\Delta}^{\perp} )p^{'+}\frac{A^{\pi^+}(x,\xi,t)}{f_{\pi}}\nonumber\\
&+e (\varepsilon\cdot \Delta) \frac{f_{\pi}}{m_{\pi}^2-t}\epsilon(\xi)\phi(\frac{x+\xi}{2\xi}),
\end{align}
\end{widetext}
where $x$ is the longitudinal momentum fraction, the pion decay constant $f_{\pi}$, $e$ is the electric charge, $\varepsilon$ is the photon polarization vector. Here, we have defined TDAs of a transition from a $\pi^+$ to a photon. Through symmetry properties, the TDAs of a $\pi^-$ to a photon involved in other processes can be obtained.
For instance, we could wish to study the $\gamma - \pi^-$ TDAs entering the factorized amplitude of the process. $V^{\pi^+}(x,\xi,t)$ and $A^{\pi^+}(x,\xi,t)$ are the vector and axial TDAs, respectively. The transverse condition is $\varepsilon \cdot p'=0$. Therefore, the axial matrix element contains the axial TDA and the pion pole contribution that has been isolated in a model independent way~\cite{Courtoy:2010qn}, which was described in Fig. \ref{pionpole}. The latter term is parameterized by a point-like pion propagator multiplied by the PDA of an on-shell pion, $\phi_{\pi}(x)$. The pion pole contribution is not considered in this paper, as the pion pole term in the t-channel is irrelevant for the evaluation of the axial TDA ~\cite{Tiburzi:2005nj}. In Ref.~\cite{Broniowski:2007fs}, the pion photon TDAs in the Spectral Quark Model were calculated without taking into account this term.
%

In the second term of Eq. (\ref{atda}), pion PDA $\phi(x)$ is introduced. The definition of pion PDA is
\begin{align}\label{atda1}
&\int \frac{\mathrm{d}z^-}{2\pi}e^{ixp^+z^-} \langle 0 |\bar{q}(-\frac{z}{2})\gamma^+\gamma^5q(\frac{z}{2})|\pi^+(p)\rangle \mid_{z^+=0,\bm{z}=\bm{0}}\nonumber\\
&=\frac{if_{\pi}}{p^+}\phi_{\pi}(x).
\end{align}

In Fig. \ref{piontda}, the insert operators $\Gamma$ look like this: 
\begin{subequations}\label{a91}
\begin{align}\label{6A}
\Gamma^V &=\gamma^+\delta(x-\frac{k^+}{P^+})\,, \\
\Gamma^A &=\gamma^+\gamma^5 \delta(x-\frac{k^+}{P^+}),
\end{align}
\end{subequations}
$\Gamma^V$ for vector TDA and $\Gamma^A$ for axial TDA.

We believe that the handbag diagram dominates the process. There are two related contributions to each TDA, depending on which quark ($u$ or $d$) of the pion is scattered off by the deep virtual photon. For an active $u$-quark in Fig. \ref{piontda}, in the NJL model, $u$-quark pion TDAs are defined as
\begin{align}\label{gpddd}
&v_{u\rightarrow d}^{\pi^+}\left(x,\xi,t\right)
=i e N_c \sqrt{Z_{\pi}}\varepsilon^{*\nu} \int \frac{\mathrm{d}^4k}{(2 \pi )^4}\delta_n^x (k)\nonumber\\
&\times \text{tr}_{\text{D}}\left[\gamma _{\nu} S_u \left(k_{+\Delta}\right)\gamma ^+ S_u\left(k_{-\Delta}\right)\gamma _5 S_d\left(k-P\right)\right],
\end{align}
\begin{align}\label{tgpddd}
& a_{u\rightarrow d}^{\pi^+}\left(x,\xi,t\right)=i e N_c \sqrt{Z_{\pi}} \varepsilon^{*\nu}  \int \frac{\mathrm{d}^4k}{(2 \pi )^4}\delta_n^x (k)\nonumber\\
&\times \text{tr}_{\text{D}}\left[\gamma _{\nu} S_u \left(k_{+\Delta}\right)\gamma^+\gamma^5 S_u\left(k_{-\Delta}\right)\gamma _5 S_d\left(k-P\right)\right],
\end{align}
\begin{align}\label{ab25}
\phi_{\pi}(x)&=\frac{iN_c p^+}{f_{\pi }} \int \frac{\mathrm{d}^4k}{(2 \pi )^4}\delta(xp^+-k^+)\nonumber\\
&\times \text{tr}_{\text{D}}\left[\gamma^+\gamma^5 S_u \left(k\right)\gamma^5 S_d\left(k-p\right)\right],
\end{align}
where $\text{tr}_{\text{D}}$ indicates a trace over spinor indices, $\delta_n^x (k)=\delta (xP^+-k^+)$, $k_{+\Delta}=k+\frac{\Delta}{2}$, $k_{-\Delta}=k-\frac{\Delta}{2}$.

After a thorough calculation, the final form of TDAs are
\begin{align}\label{agpdf}
v_{u\rightarrow d}^{\pi^+}\left(x,\xi,t\right)=\frac{ N_cf_{\pi}\sqrt{Z_{\pi }} }{4\pi ^2}\int_0^1 \mathrm{d}\alpha \frac{\theta_{\alpha \xi}}{\xi} M_u \frac{\bar{\mathcal{C}}_2(\sigma_2)}{\sigma_2},
\end{align}
\begin{align}\label{agtpdf}
a_{u\rightarrow d}^{\pi^+}\left(x,\xi,t\right)&=-\frac{N_cf_{\pi}\sqrt{Z_{\pi } }M_u}{4\pi ^2}\int_0^1 \mathrm{d}\alpha \frac{\theta_{\alpha \xi}}{\xi}\nonumber\\
&\times \left(\frac{x}{\xi }-\frac{\alpha  (1-\xi )}{\xi }\right)\frac{\bar{\mathcal{C}}_2(\sigma_2)}{\sigma_2},
\end{align}
and
\begin{align}\label{thetaf}
\theta_{\alpha \xi}&=x\in[\alpha (\xi +1)-\xi , \alpha  (1-\xi)+\xi ]\cap x\in[-1,1],
\end{align}
where $x$ only exist in the corresponding region. One can write $\theta_{\bar{\xi} \xi}/\xi=\Theta(1-x^2/\xi^2) $, where $\Theta(x)$ is the Heaviside function, and $\theta_{\alpha \xi}/\xi=\Theta((1-\alpha^2)-(x-\alpha)^2/\xi^2)\Theta(1-x^2)$. These results are in the region $\xi > 0$. Unlike the pion GPDs~\cite{Zhang:2021shm}, TDAs no longer have symmetry properties in $x$ and $\xi$.

The pion PDA in the NJL model is
\begin{align}\label{ab25}
\phi_{\pi}(x)=\frac{3 M_u \sqrt{Z_{\pi }}}{4 \pi ^2 f_{\pi }}\bar{\mathcal{C}}_1(\sigma_1),
\end{align}
PDA is defined within the region $x\in [0,1]$ and meets the condition
\begin{align}\label{ab25}
\int_0^1  \phi_{\pi}(x) \mathrm{d}x=1,
\end{align}
in the chiral limit $m_{\pi}=0$, $\phi_{\pi}(x)=1$, when $m_{\pi}=0.14$ GeV, $\phi_{\pi}(x)$ just varies a little near $1$. 

One can obtain
\begin{align}\label{ab25}
\phi_{\pi}(\frac{x+\xi}{2\xi})=\frac{3 M_u \sqrt{Z_{\pi }}}{4 \pi ^2 f_{\pi }}\int_{\tau_{uv}^2}^{\tau_{ir}^2} d\tau \frac{1}{\tau }e^{-\tau (M_u^2+\frac{x^2-\xi^2}{4\xi^2}m_{\pi}^2)},
\end{align}
where $x\in[-\xi,\xi]$, that is the ERBL region, the kinematics in this region enable the emission or absorption of a pion from the initial state.

%
%

In the following section, we will check the basic properties of TDAs.

\subsection{The properties of pion photon TDAs}\label{qq}
\subsubsection{Isospin relations}
\begin{figure*}
\centering
\includegraphics[width=0.47\textwidth]{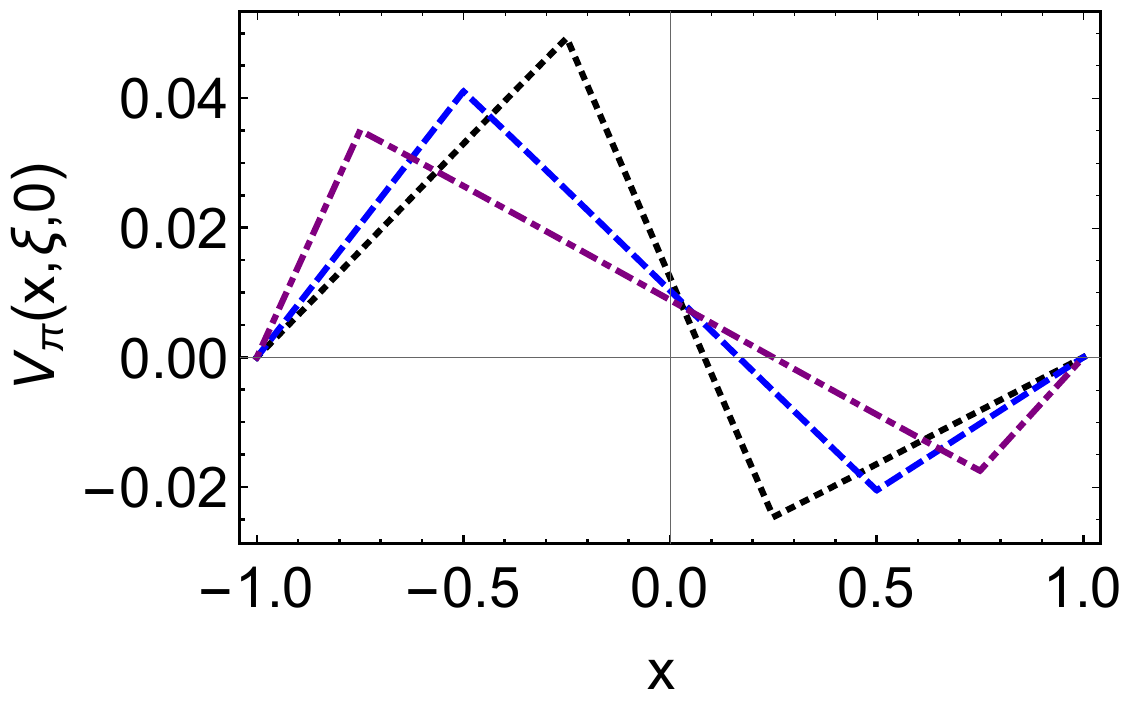}
\qquad
\includegraphics[width=0.47\textwidth]{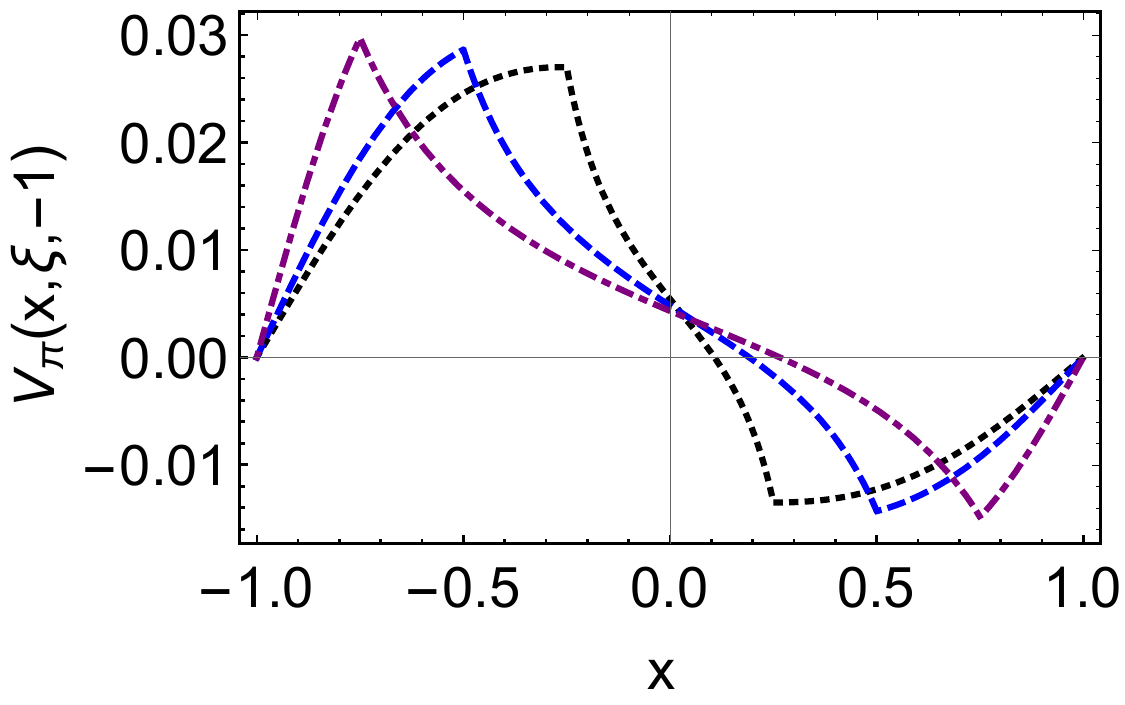}
\caption{The pion-photon vector TDAs (left panel: $V^{\pi^+}(x,\xi,0)$, right panel: $V^{\pi^+}(x,\xi,-1)$): $\xi=0.25$ --- black dotted curve, $\xi=0.5$ --- blue dashed curve, $\xi=0.75$ --- purple dotdashed curve. }\label{pitda}
\end{figure*}

\begin{figure*}
\centering
\includegraphics[width=0.47\textwidth]{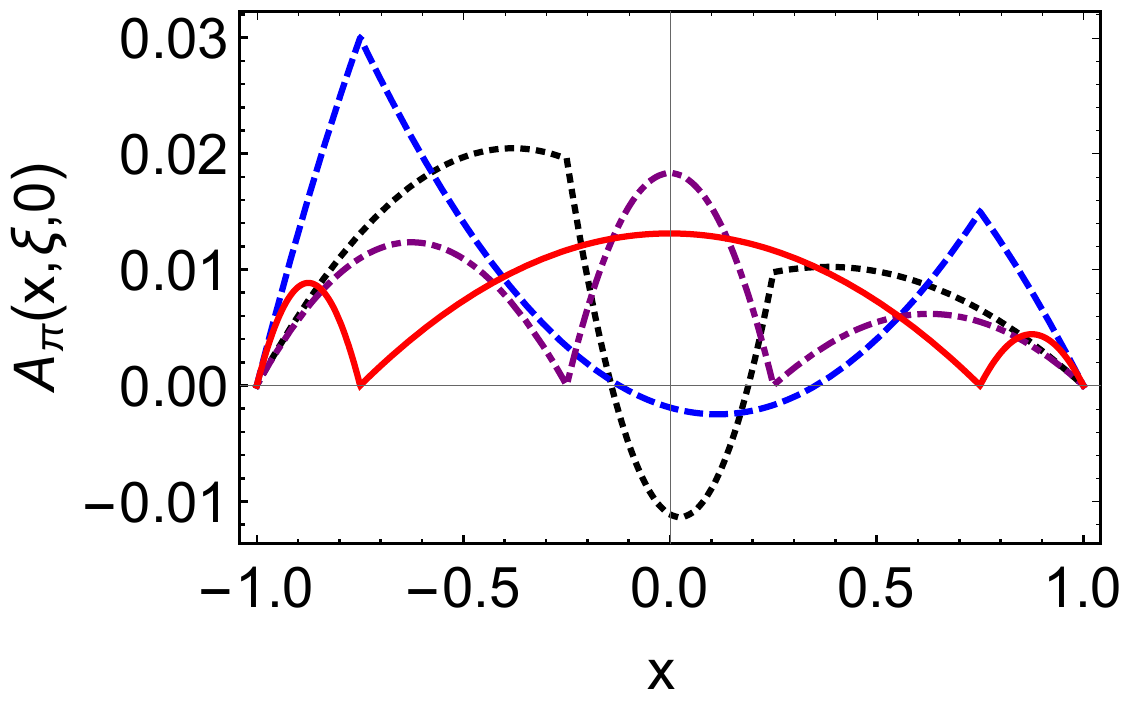}
\qquad
\includegraphics[width=0.47\textwidth]{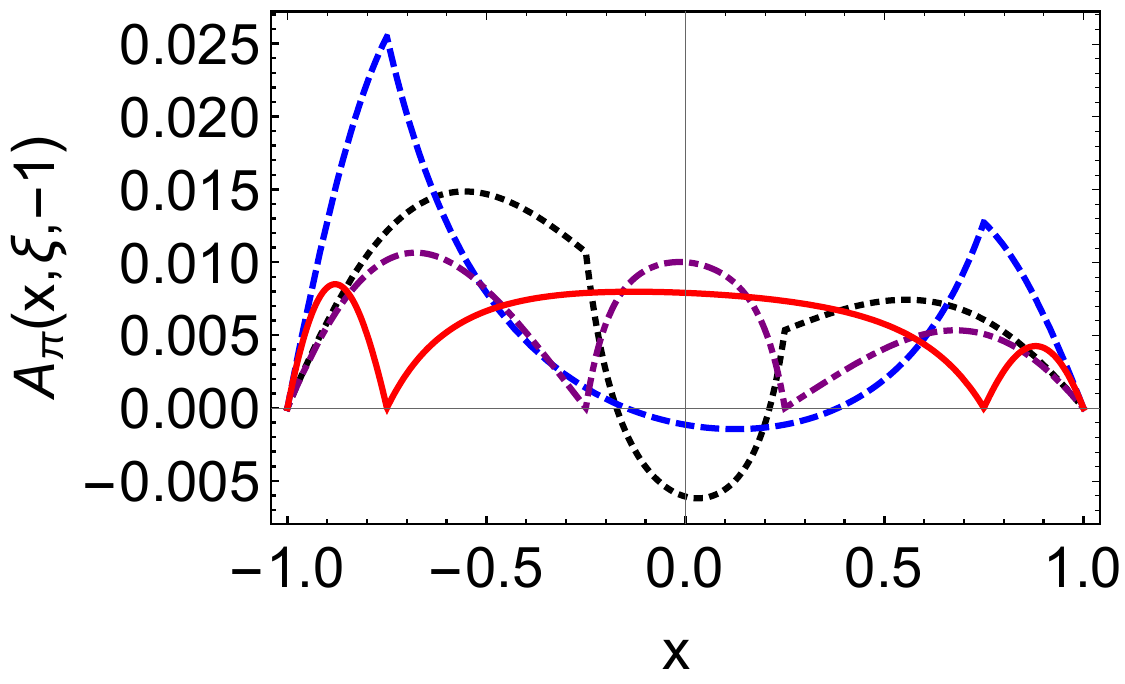}
\caption{The pion-photon axial vector TDAs (left panel: $A^{\pi^+}(x,\xi,0)$, right panel: $A^{\pi^+}(x,\xi,-1)$): $\xi=0.25$ --- black dotted curve, $\xi=0.75$ --- blue dashed curve, $\xi=-0.25$ --- purple dotdashed curve, $\xi=-0.75$ --- red solid curve. }\label{pitda1}
\end{figure*}

Isospin relates the two contributions,
\begin{subequations}
\begin{align}\label{ab94}
v_{\bar{d}\rightarrow \bar{u}}^{\pi^+}\left(x,\xi,t\right)=v_{u\rightarrow d}^{\pi^+}\left(-x,\xi,t\right) \,, \\
a_{\bar{d}\rightarrow \bar{u}}^{\pi^+}\left(x,\xi,t\right)=-a_{u\rightarrow d}^{\pi^+}\left(-x,\xi,t\right),
\end{align}
\end{subequations}
then we can obtain
\begin{subequations}\label{com}
\begin{align}
V^{\pi^+}(x,\xi,t) &= Q_d v_{u\rightarrow d}^{\pi^+} (x,\xi,t)+Q_uv_{\bar{d}\rightarrow \bar{u}}^{\pi^+}\left(x,\xi,t\right)\,, \\
A^{\pi^+}(x,\xi,t)&= Q_d a_{u\rightarrow d}^{\pi^+} (x,\xi,t)+Q_ua_{\bar{d}\rightarrow \bar{u}}^{\pi^+}\left(x,\xi,t\right),
\end{align}
\end{subequations}
the diagrams of vector and axial vector TDAs have been plotted separately in Figs. \ref{pitda} and \ref{pitda1}. We only choose three cases for the vector TDA when $\xi$ is positive because the behaviors of the negative $\xi$ are the same. This means that it is an even function of the skewness variable. The axial TDAs behave differently because of the sign of $xi$, compared to the vector TDAs. The maxima value for the negative $\xi$ is around $x=0$, whereas for the positive $\xi$, the contribution exhibits the opposite sign. From Eqs. (\ref{com}) we obtain that in the region $|\xi| \leq x \leq 1$ and $-1 \leq x \leq -|\xi|$, the isospin relates the values of the vector and the axial vector TDAs,
\begin{subequations}\label{com1}
\begin{align}
V(x,\xi,t) &= -\frac{1}{2}V(-x,\xi,t)\,, \\
A(x,\xi,t) &= \frac{1}{2}A(-x,\xi,t),
\end{align}
\end{subequations}
where $1/2$ comes from the ratio of charge between $u$ and $d$ quark. Our results matches these relationships, and the relationship will not change by evolution. The figures are similar to the Refs. ~\cite{Courtoy:2007vy,Courtoy:2008af,Courtoy:2008ij,Kotko:2008gy,Kotko:2013ima}. Axial and vector TDAs in the NJL model are calculated  using the Pauli-Villars regularization procedure in Ref.~\cite{Courtoy:2007vy}. When compared to their diagrams, it can be observed that the TDAs are qualitatively similar but quantitatively different, indicating that regularization has a negligible impact on the outcomes of the TDAs. In Ref.~\cite{Kotko:2008gy}, they study the TDAs in the non-local chiral quark model, which is different from the vector TDA in the local NJL model. Vector TDA is no longer symmetric in the non-local case. The behaviors of axial TDAs are similar in both the local and non-local model for the positive $\xi$. For the negative $\xi$, the differences are that at the points $x=\pm \xi$, in the local model, the values of axial TDAs are zero, but not in non-local model.




Time reversal relates the $\pi^+-\gamma$ TDAs to $\gamma-\pi^+$ TDAs in the following way
\begin{align}\label{ab94}
D^{\pi^+-\gamma}(x,\xi,t) = D^{\gamma-\pi^+}(x,-\xi,t)
\end{align}
where D represents the vector and axial vector TDAs. CPT relates the presently calculated TDAs to their analog for a transition from a photon to a $\pi^-$
\begin{subequations}
\begin{align}\label{ab94}
V^{\pi^+-\gamma}(x,\xi,t)&= V^{\gamma-\pi^-}(-x,-\xi,t) \,, \\
A^{\pi^+-\gamma}(x,\xi,t)&= -A^{\gamma-\pi^-}(-x,-\xi,t).
\end{align}
\end{subequations}


\subsubsection{Form factors}
The pion TDAs satisfy the sum rule
\begin{subequations}
\begin{align}\label{ab94}
\int_{-1}^1 V^{\pi^+}(x,\xi,t) \mathrm{d}x&= \frac{ f_{\pi} }{m_{\pi}} F_V^{\pi^+}(t)\,, \\
\int_{-1}^1 A^{\pi^+} (x,\xi,t) \mathrm{d}x&= \frac{ f_{\pi} }{m_{\pi}} F_A^{\pi^+}(t),
\end{align}
\end{subequations}
then we can obtain
\begin{align}\label{aF3}
F_V^{\pi^+}(t)&=\frac{\sqrt{Z_{\pi }}}{2\pi^2} \int_0^1 \mathrm{d}x \int_0^{1-x}\mathrm{d}y\frac{M_u m_{\pi}}{\sigma_3}\bar{\mathcal{C}}_2(\sigma_3),
\end{align}
\begin{align}\label{a1F3}
F_A^{\pi^+}(t)&=\frac{ N_c\sqrt{Z_{\pi }} }{2\pi ^2}  \int _0^1\mathrm{d}x \int _0^{1-x}\mathrm{d}y \frac{M_u m_{\pi}(1-2y)}{\sigma_3}\bar{\mathcal{C}}_2(\sigma_3).
\end{align}
The two transition FFs are identical to the results computed from the definition of the transition FFs in the NJL model, indicating that our TDAs satisfy the sum rule constraint. One can obtain $F_V^{\pi^+}(0)=0.0234$, which is in agreement with the experimental value $F_V(0) = 0.017 \pm 0.008$ given in Ref.~\cite{ParticleDataGroup:2006fqo}. We obtain the axial TDA at $t=0$, $F_A^{\pi^+}(0)=0.0233$, the prediction value by PDG is $F_A^{\pi^+}(0)=0.0115 \pm 0.0005$~\cite{Pire:2004ie}. The estimate ratio of $F_A^{\pi^+}(0)/F_V^{\pi^+}(0)$ is $0.7$, our results give $F_A^{\pi^+}(0)/F_V^{\pi^+}(0)\approx 1$, which is similar to the results of Refs.~\cite{Courtoy:2007vy,Broniowski:2007fs}.

\begin{figure*}
\centering
\includegraphics[width=0.47\textwidth]{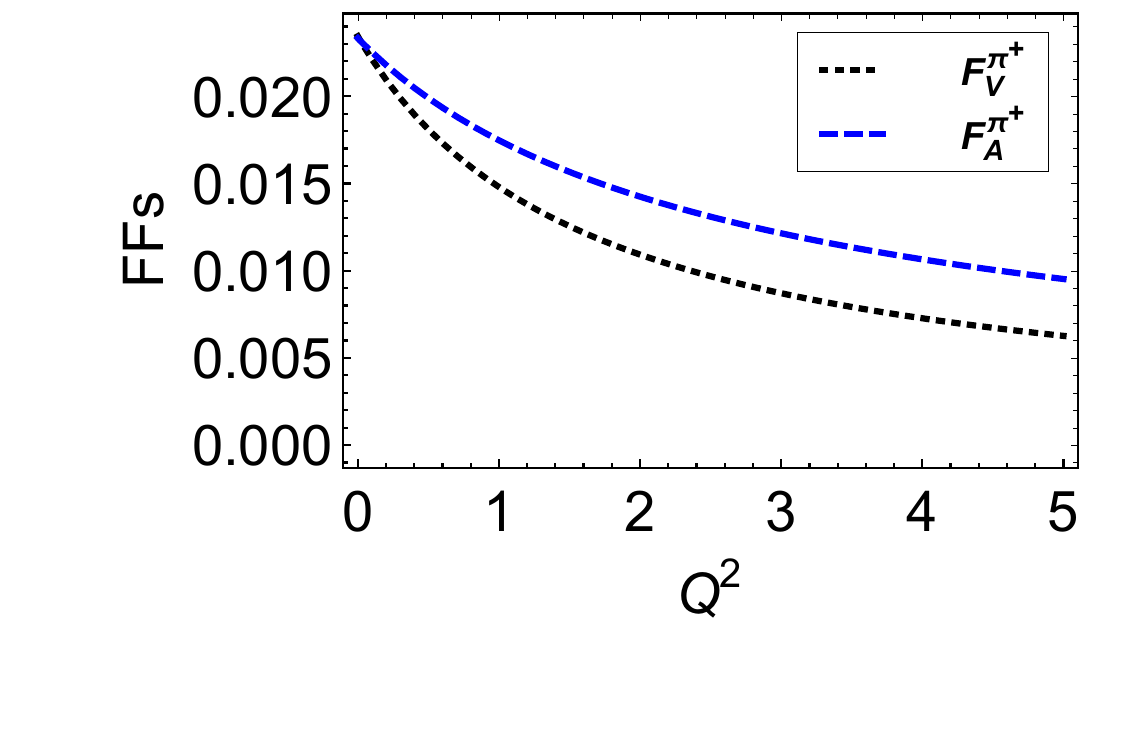}
\qquad
\includegraphics[width=0.47\textwidth]{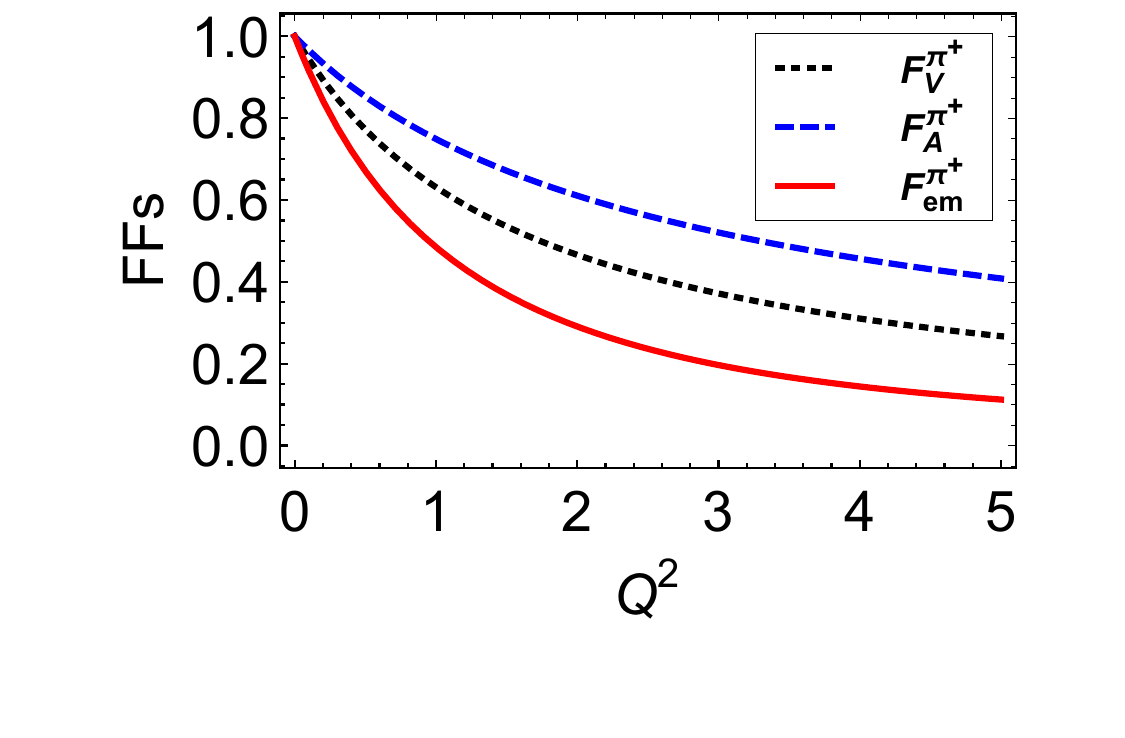}
\caption{The pion transition FFs, left panel: vector FFs $F_V^{\pi^+}(Q^2)$ -- dotted black curve; axial FF $F_A^{\pi^+}(Q^2)$ -- dashed black curve; right panel: $F_V^{\pi^+}(Q^2)/F_V^{\pi^+}(0)$ -- dotted black curve; $F_A^{\pi^+}(Q^2)/F_A^{\pi^+}(0)$ -- dashed black curve, electromagnetic FF $F_{em}^{\pi^+}(Q^2)/F_{em}^{\pi^+}(0)$ -- red solid curve.}\label{piff}
\end{figure*}

The $\pi^0$ distribution must satisfy the following sum rule~\cite{Pire:2004ie}
\begin{align}\label{ab94}
\int_{-1}^1 \mathrm{d}x(Q_u V_u^{\pi^0}(x,\xi,t)-Q_dV_u^{\pi^0}(x,\xi,t)) =\sqrt{2}f_{\pi}F_{\pi \gamma^*\gamma}(t)
\end{align}
$F_{\pi \gamma^*\gamma}$ is the pion photon transition form factor, which is directly related to $F_V$. Ref.~\cite{Brodsky:1981rp} gives a theoretical prediction for the $\pi^0$ form factor $F_{\pi \gamma^*\gamma}(t)$, at $t=0$, $F_{\pi \gamma^*\gamma}(0)=0.272$ GeV$^{-1}$. The neutral pion FF depends on the sum rule of $\pi^+$ vector FF. Our results give $F_{\pi \gamma^*\gamma}(0)=0.237$ GeV$^{-1}$, we have plot the pion photon transition FFs in Figs. \ref{piff}.

\subsubsection{Polynomiality condition}
The pion TDAs should satisfy the polynomiality condition, for the vector TDA
\begin{align}\label{ab94}
\int_{-1}^1 x^{n-1} V^{\pi^+}(x,\xi,t) \mathrm{d}x= \sum_{i=0}^{n-1}C_{n,i}(t)\xi^i,
\end{align}
which means the polynomials include all powers of $\xi$. But in the chiral limit $m_{\pi}^2=0$, there are only the even powers of $\xi$ left in the polynomial expansion, this is consistent with Ref.~\cite{Courtoy:2007vy}.

For the $\xi$-dependence of the moments of axial TDA
\begin{align}\label{ab94}
\int_{-1}^1 x^{n-1} A^{\pi^+}(x,\xi,t) \mathrm{d}x= \sum_{i=0}^{n-1}C_{n,i}^{'}(t)\xi^i,
\end{align}
similar to the vector TDA, the axial TDA contains all the powers of $\xi$. Different from the vector TDA, the polynomials of axial TDA still contain all the powers of $\xi$ in the chiral limit.

\subsection{The properties of kaon-photon TDAs}\label{qq}
The results of kaon photon TDAs in the NJL are as follows
\begin{widetext}
\begin{align}\label{agpdf}
v_{u\rightarrow s}^{K^+}\left(x,\xi,t\right)&=-\frac{ N_c f_K\sqrt{Z_K} }{4\pi ^2}\int_0^1 \mathrm{d}\alpha \frac{\theta_{\alpha \xi}}{\xi}((1-\alpha ) \left(M_u-M_s\right)+M_s) \frac{\bar{\mathcal{C}}_2(\sigma_5)}{\sigma_5},
\end{align}
\begin{align}\label{agtpdf}
&a_{u\rightarrow s}^{K^+}\left(x,\xi,t\right)=-\frac{N_c f_K\sqrt{Z_K }}{4\pi ^2}\int_0^1 \mathrm{d}\alpha \frac{\theta_{\alpha \xi}}{\xi}\left(M_u-M_s\right) \left(\frac{\alpha(2\alpha -1) (1-\xi )}{2\xi }+\frac{(1-\alpha ) x}{\xi }-\frac{x-\xi }{2 \xi }\right)\frac{\bar{\mathcal{C}}_2(\sigma_5)}{\sigma_5}\nonumber\\
&-\frac{N_cf_K\sqrt{Z_K }}{4\pi ^2}\int_0^1 \mathrm{d}\alpha \frac{\theta_{\alpha \xi}}{\xi} \left(\left(M_s+M_u\right) \left(\frac{x-\xi }{2 \xi }-\frac{\alpha  (1-\xi )}{2 \xi }\right)+M_s \right)\frac{\bar{\mathcal{C}}_2(\sigma_5)}{\sigma_5},
\end{align}
\end{widetext}
where $\theta_{\alpha \xi}$ is defined in Eq. (\ref{thetaf}).

In the NJL model, the kaon PDA is
\begin{align}\label{ab25}
\phi_{K}(x)=\frac{3 \sqrt{Z_K}}{4 \pi ^2 f_K}( M_u +x(M_s-M_u)) \bar{\mathcal{C}}_1(\sigma_4),
\end{align}
PDA is defined in the region $x\in [0,1]$ and satisfies the condition
\begin{align}\label{ab25}
\int_0^1  \phi_{K}(x) \mathrm{d}x=1,
\end{align}
different from pion PDA $\phi_{\pi}(x)$, kaon PDA $\phi_K(x)$ is no longer symmetry at $x=1/2$.

Then we obtain
\begin{align}\label{ab25}
\phi_{K}(\frac{x+\xi}{2\xi})&=\frac{3\sqrt{Z_K}}{4 \pi ^2 f_K}\int_{\tau_{uv}^2}^{\tau_{ir}^2} d\tau \frac{1}{\tau }( \frac{\xi-x}{2\xi}M_u +\frac{x+\xi}{2\xi}M_s)\nonumber\\
&\times e^{-\tau (\frac{\xi-x}{2\xi}M_u^2+\frac{x+\xi}{2\xi}M_s^2+\frac{x^2-\xi^2}{4\xi^2}m_K^2)},
\end{align}
where $x$ is also in the ERBL region $x\in[-\xi,\xi]$.

\subsubsection{Isospin relations}
\begin{figure*}
\centering
\includegraphics[width=0.47\textwidth]{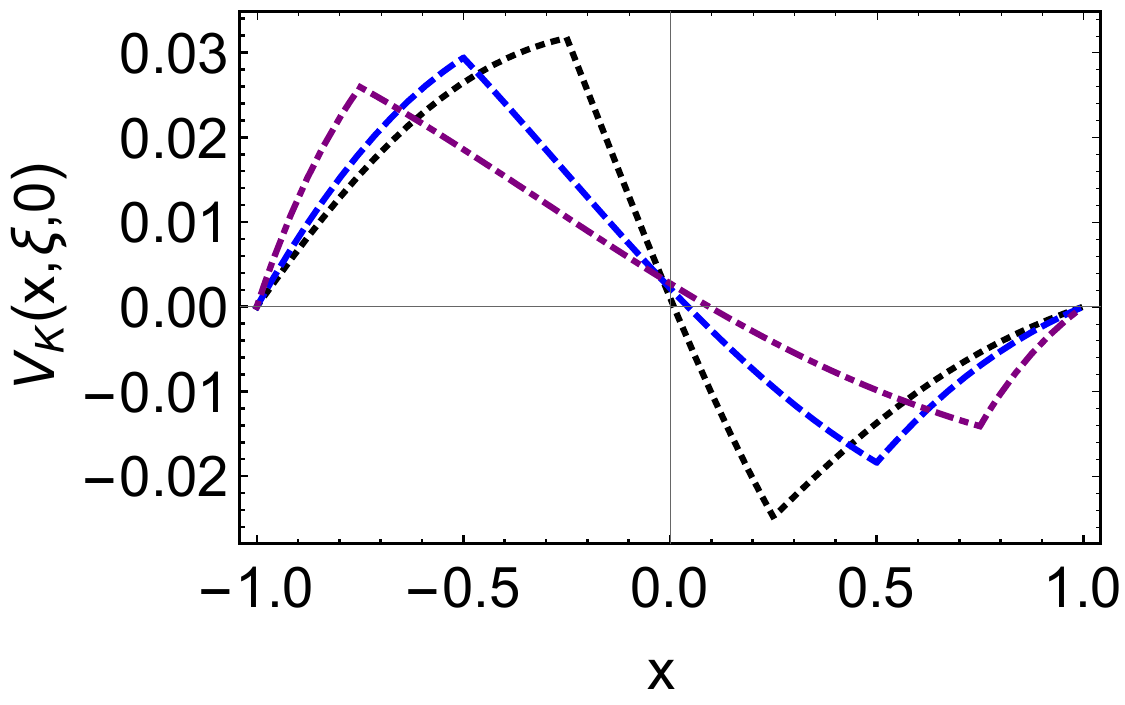}
\qquad
\includegraphics[width=0.47\textwidth]{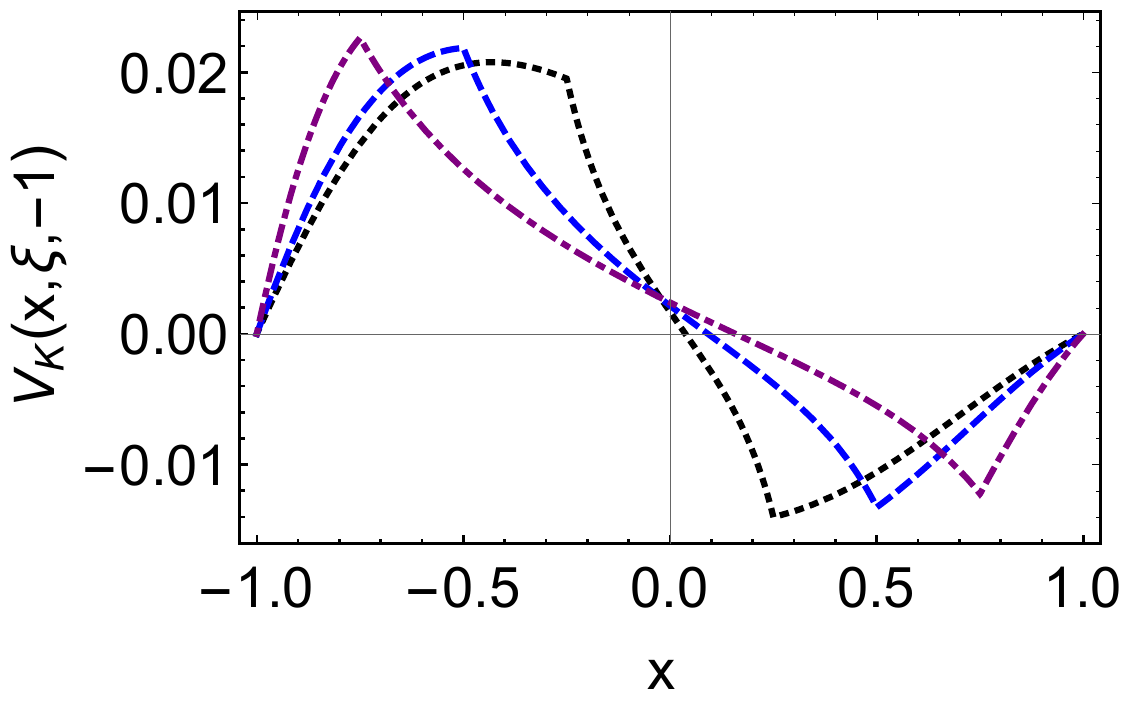}
\caption{The kaon-photon vector TDAs (left panel: $V^{K^+}(x,\xi,0)$, right panel: $V^{K^+}(x,\xi,-1)$): $\xi=0.25$ --- black dotted curve, $\xi=0.5$ --- blue dashed curve, $\xi=0.75$ --- purple dotdashed curve. }\label{katda}
\end{figure*}

\begin{figure*}
\centering
\includegraphics[width=0.47\textwidth]{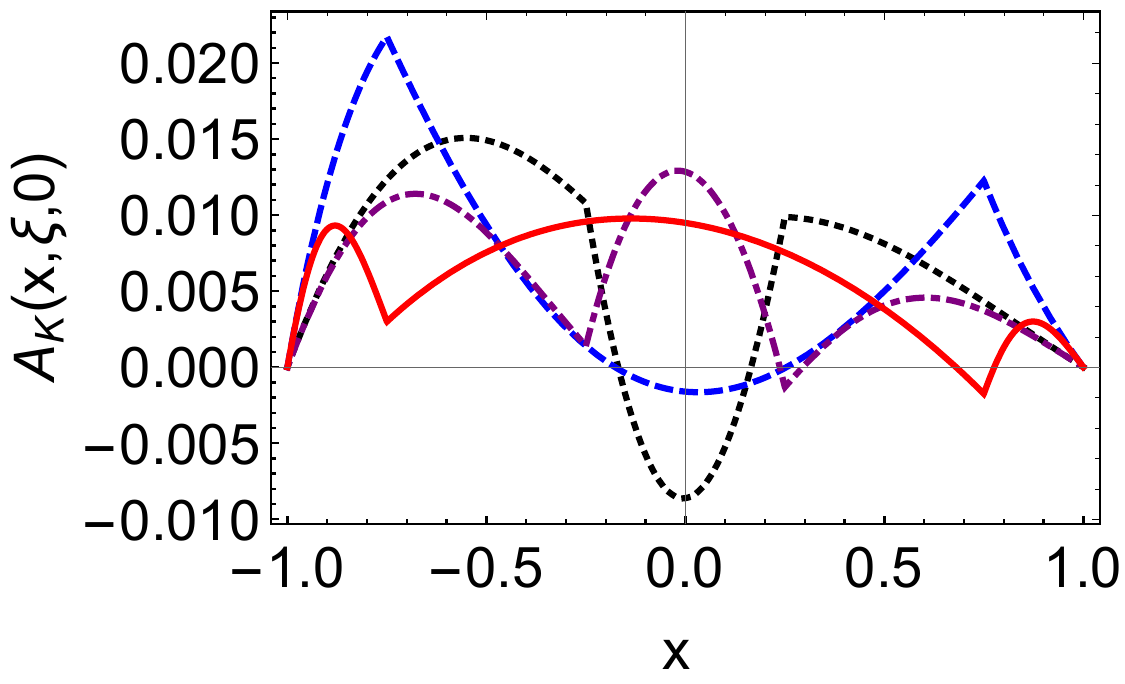}
\qquad
\includegraphics[width=0.47\textwidth]{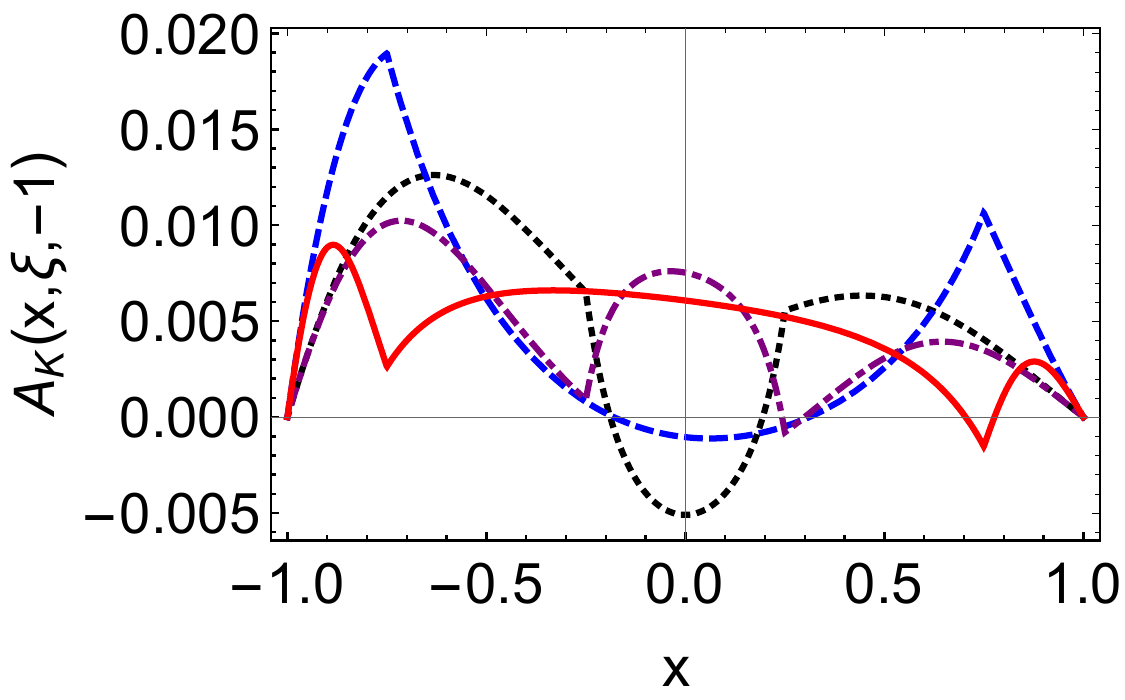}
\caption{The kaon-photon axial vector TDAs (left panel: $A^{K^+}(x,\xi,0)$, right panel: $A^{K^+}(x,\xi,-1)$): $\xi=0.25$ --- black dotted curve, $\xi=0.75$ --- blue dashed curve, $\xi=-0.25$ --- purple dotdashed curve, $\xi=-0.75$ --- red solid curve. }\label{katda1}
\end{figure*}

Isospin relates these two contributions,
\begin{widetext}
\begin{subequations}
\begin{align}\label{ab94}
v_{\bar{s}\rightarrow \bar{u}}^{K^+}\left(x,\xi,t,M_u,M_s\right)=v_{u\rightarrow s}^{K^+}\left(-x,\xi,t,M_u\leftrightarrow M_s,M_s\leftrightarrow  M_u\right) \,, \\
a_{\bar{s}\rightarrow \bar{u}}^{K^+}\left(x,\xi,t,M_u,M_s\right)=-a_{u\rightarrow s}^{K^+}\left(-x,\xi,t,M_u\leftrightarrow M_s,M_s\leftrightarrow  M_u\right),
\end{align}
\end{subequations}
\end{widetext}
then we can obtain
\begin{subequations}
\begin{align}\label{ab94}
V^{K^+}(x,\xi,t) &= Q_s v_{u\rightarrow s}^{K^+} (x,\xi,t)+Q_uv_{\bar{s}\rightarrow \bar{u}}^{K^+}\left(x,\xi,t\right)\,, \\
A^{K^+}(x,\xi,t)&= Q_s a_{u\rightarrow s}^{K^+} (x,\xi,t)+Q_ua_{\bar{s}\rightarrow \bar{u}}^{K^+}\left(x,\xi,t\right),
\end{align}
\end{subequations}
we have plotted the diagrams of kaon vector and axial vector TDAs in Figs. \ref{katda} and \ref{katda1} separately. As shown in the diagram, the kaon TDAs are similar to pion TDAs. For the axial TDAs, when $\xi$ is negative, the difference is that at the points $x= \pm \xi $, pion photon TDAs are zero, but kaon photon TDAs are not, the values at $x=\xi$ are positive, the values at $x=-\xi$ are negative. The first thing to note is that the relations in Eq. (\ref{com1}) are invalid for kaon TDAs, which is due to the breaking of isospin symmetry. Secondly, for the vector TDAs, the maximum value of pion is larger than the kaon vector TDA, and the quality behavior is similar. The kaon axial TDA behavior is similar to that of pion results when $\xi$ is positive. When $\xi$ is negative, the values are very different at the points $x=\pm\xi$, the values are not zero anymore. The values are positive at $x=\xi$, but negative at $x=-\xi$.


\subsubsection{Form factors}

\begin{figure*}
\centering
\includegraphics[width=0.47\textwidth]{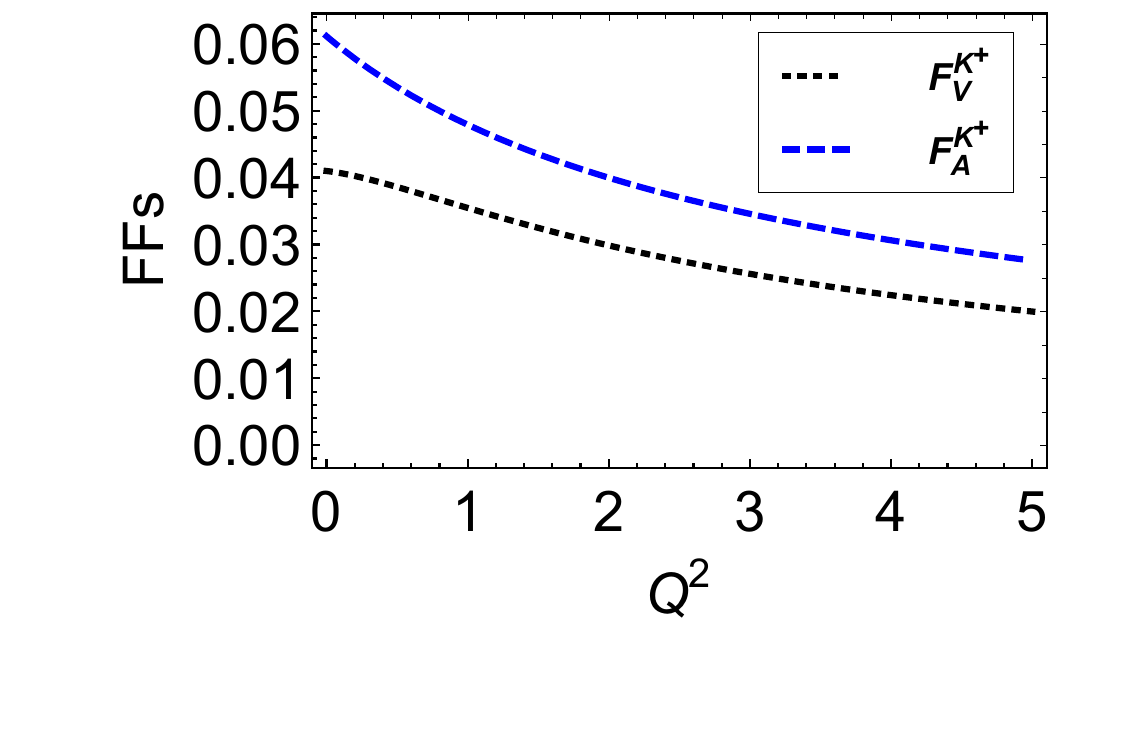}
\qquad
\includegraphics[width=0.47\textwidth]{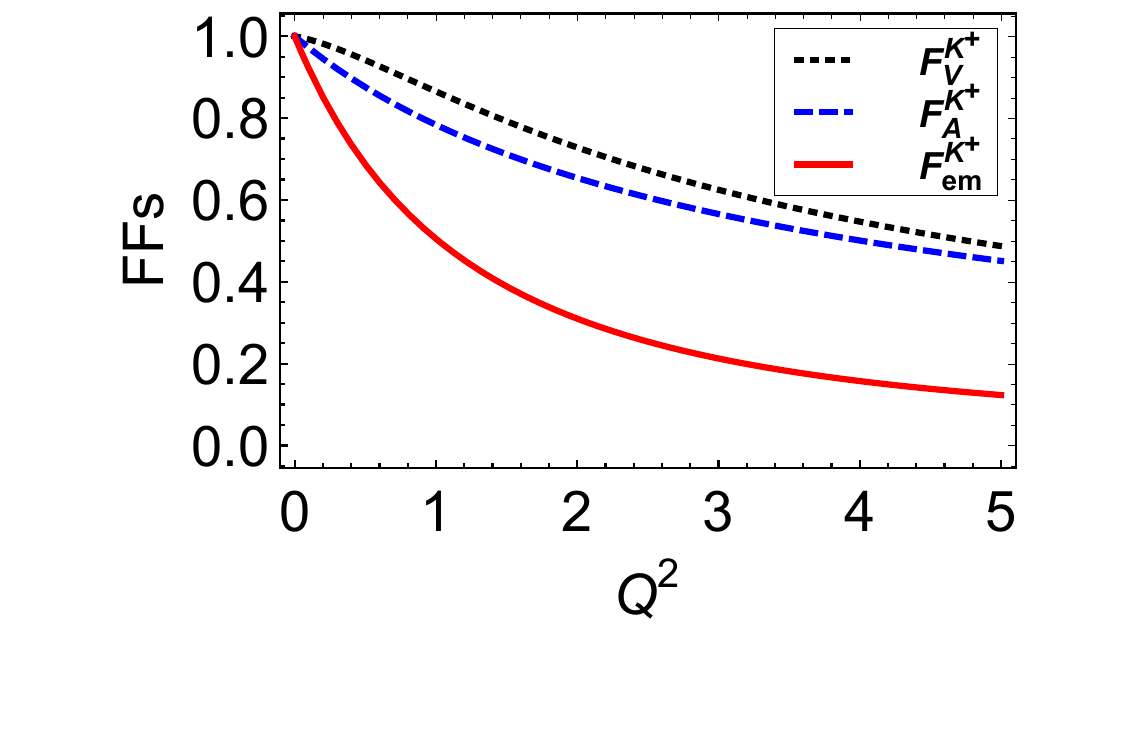}
\caption{The kaon transition FFs, left panel: vector FFs $F_V^{K^+}(Q^2)$ -- dotted black curve; axial FF $F_A^{K^+}(Q^2)$ -- dashed black curve; right panel: $F_V^{K^+}(Q^2)/F_V^{K^+}(0)$ -- dotted black curve; $F_A^{K^+}(Q^2)/F_A^{K^+}(0)$ -- dashed black curve, electromagnetic FF $F_{em}^{K^+}(Q^2)/F_{em}^{K^+}(0)$ -- red solid curve.}\label{kaff}
\end{figure*}

The kaon vector and axial form factors
\begin{widetext}
\begin{align}\label{aF3}
F_V^{K^+}(t)&=-\frac{\sqrt{Z_K}}{2\pi^2} \int_0^1 \mathrm{d}x \int_0^{1-x}\mathrm{d}y\,\frac{\bar{\mathcal{C}}_2(\sigma_6)}{\sigma_6} m_K \left(M_s+(x+y) \left(M_u-M_s\right)\right)\nonumber\\
&+\frac{\sqrt{Z_K}}{\pi^2} \int_0^1 \mathrm{d}x \int_0^{1-x}\mathrm{d}y\,\frac{\bar{\mathcal{C}}_2(\sigma_7)}{\sigma_7} m_K \left(M_u-(x+y) \left(M_u-M_s\right)\right),
\end{align}
\begin{align}\label{a1F3}
F_A^{K^+}(t)&=\frac{ \sqrt{Z_K} }{2\pi ^2}  \int _0^1\mathrm{d}x \int _0^{1-x}\mathrm{d}y \frac{\bar{\mathcal{C}}_2(\sigma_6)}{\sigma_6} m_K \left((2 y (1-x-y)+x) \left(M_u-M_s\right)-y \left(M_s+M_u\right)+M_s\right)\nonumber\\
&+\frac{ \sqrt{Z_K} }{\pi ^2}  \int _0^1\mathrm{d}x \int _0^{1-x}\mathrm{d}y \frac{\bar{\mathcal{C}}_2(\sigma_7)}{\sigma_7} m_K \left((2 y (1-x-y)+x) \left(M_s-M_u\right)-y \left(M_s+M_u\right)+M_u\right).
\end{align}
\end{widetext}
The results are the same as the kaon photon transition form factors calculated from the definition. This implies that pion and kaon photon TDAs all satisfy the sum rule constraint. Similar to the pion TDAs, kaon TDAs also satisfy the sum rule and polynomiality condition. We plot the vector and axial FFs of kaon in Fig. \ref{kaff}. $F_V^{K^+}(0)=0.041$ GeV$^{-1}$, $F_A^{K^+}(0)=0.061$ GeV$^{-1}$. Based on the figures, it can be observed that $F_V^{K^+}(Q^2)$ is quite different from pion $F_V^{\pi^+}(Q^2)$. Kaon vector FF is harder than axial FF, that's different from pion FFs, the same point is that both of the two transition FFs are harder than electromagnetic FFs.

\section{Summary and conclusions}\label{excellent}
In the present paper, we calculate the pion-photon and kaon-photon vector and axial vector TDAs in the NJL model using the PTR scheme with an infrared cutoff to mimic confinement.

Firstly, we study the pion-photon TDAs, the diagrams of vector and axial vector TDA are plotted in Figs. \ref{pitda} and \ref{pitda1}, and the figures of vector and axial transition form factors. We have obtained a value of $F_V^{\pi^+}(0)=0.0234$, which is in agreement with the experimental value of $F_V(0) = 0.017 \pm 0.008$. The axial TDA at $t=0$, $F_A^{\pi^+}(0)=0.0233$, and the prediction value by PDG is $F_A^{\pi^+}(0)=0.0115 \pm 0.0005$. The results are similar to Ref.~\cite{Lansberg:2006fv,Courtoy:2008nf}

Secondly, we conduct an investigation of the kaon-photon TDAs by employing the identical approach in the NJL model. The TDAs are plotted in Fig. \ref{katda} and \ref{katda1}. The axial TDA at $t=0$ is $F_A^{K^+}(0)=0.0233$, the prediction value by PDG is $F_A^{K^+}(0)=0.0115 \pm 0.0005$. The pion and kaon photon transition form factors are compared in Fig. \ref{kaff}, and we can see that, unlike the pion case, the vector transition form factor is harder than the axial form factor, what they also share is that electromagnetic form factors are always the softest. The difference between the axial TDAs is that at the points $x= \pm \xi $, pion photon TDAs are zero, but kaon photon TDAs are not. At $x=\xi$ the value is positive, at $x=-\xi$ the value is negative. Both the pion and kaon TDAs are shown to satisfy the sum rule and polynomiality condition.

In summary, we have provided an explicit calculation of pion-photon and kaon-photon vector and axial leading-twist TDAs in the NJL model, and compare the difference between them. The research on TDAs gives rise to interesting evaluations of cross-sections for exclusive meson pair production in $\gamma\gamma^*$ scattering. The
analytical results provide insight into the possible shapes of non-perturbatively generated TDAs. Such estimation for TDAs or GPDs are useful, they are constraints by form factors, polynomiality, etc. Our results correspond to a low-energy scale. After suitable QCD evolution, the acquired results may be used in the research of the virtual Compton scattering and other exclusive processes involving pion and photons.


\acknowledgments
Work supported by: the Scientific Research Foundation of Nanjing Institute of Technology (Grant No. YKJ202352), the Natural Science Foundation of Jiangsu Province (Grant No. BK20191472), and the China Postdoctoral Science Foundation (Grant No. 2022M721564).




\appendix
\section{Appendix 1: useful formulas}\label{AppendixT1}

Here we use the gamma-functions ($n\in \mathbb{Z}$, $n\geq 0$)
\begin{subequations}\label{cfun}
\begin{align}
\mathcal{C}_0(z)&:=\int_0^{\infty} \mathrm{d}s\, s \int_{\tau_{uv}^2}^{\tau_{ir}^2} \mathrm{d}\tau \, e^{-\tau (s+z)}\nonumber\\
&=z[\Gamma (-1,z\tau_{uv}^2 )-\Gamma (-1,z\tau_{ir}^2 )]\,, \\
\mathcal{C}_n(z)&:=(-)^n\frac{z^n}{n!}\frac{\mathrm{d}^n}{\mathrm{d}\sigma^n}\mathcal{C}_0(z)\,, \\
\bar{\mathcal{C}}_i(z)&:=\frac{1}{z}\mathcal{C}_i(z),
\end{align}
\end{subequations}
where $\tau_{uv,ir}=1/\Lambda_{\text{UV},\text{IR}}$ are, respectively, the infrared and ultraviolet regulators described above, with $\Gamma (\alpha,y )$ being the incomplete gamma-function, $z$ represent the $\sigma$ functions in the following.

The $\sigma$ functions are define as
\begin{subequations}\label{cfun1}
\begin{align}
\sigma_1&=M_u^2-x(1-x)m_{\pi}^2\,, \\
\sigma_2&=M_u^2-\alpha \left(\frac{\xi-x}{2\xi}+\alpha \frac{1-\xi}{2\xi}\right) m_{\pi}^2\nonumber\\
&-\left(\frac{\xi+x}{2\xi}-\alpha \frac{1+\xi}{2\xi}\right) \left(\frac{\xi-x}{2\xi}+\alpha \frac{1-\xi}{2\xi}\right) t\,, \\
\sigma_3&=y(x+y-1)m_{\pi }^2-xyt+M_u^2\,, \\
\sigma_4&=(1-x)M_u^2+xM_s^2-x(1-x)m_K^2\,, \\
\sigma_5&=(1-\alpha)M_u^2+\alpha M_s^2-\alpha  \left(\frac{\xi-x}{2\xi}+\alpha \frac{1-\xi}{2\xi}\right)m_K^2\nonumber\\
&-\left(\frac{\xi+x}{2\xi}-\alpha \frac{1+\xi}{2\xi}\right) (\frac{\xi-x}{2\xi}+\alpha \frac{1-\xi}{2\xi}) t\,, \\
\sigma_6&=y(x+y-1)m_K^2-xyt\nonumber\\
&+(x+y)M_u^2+(1-x-y)M_s^2\,, \\
\sigma_7&=y(x+y-1)m_K^2-xyt\nonumber\\
&+(x+y)M_s^2+(1-x-y)M_u^2.
\end{align}
\end{subequations}

\bibliographystyle{apsrev4-1}
\bibliography{zhang}


\end{document}